\documentclass[12pt]{iopart}

\usepackage{graphicx}
\newcommand{\Vec}[1]{\mbox{\boldmath$#1$}}
\begin{document}

\title[]{First Principles 
Study on the origin of large thermopower in hole doped LaRhO$_3$ and CuRhO$_2$}

\author{Hidetomo Usui$^1$, Ryotaro Arita$^2$, Kazuhiko Kuroki$^1$}

\address{$^1$Department of Applied Physics and Chemistry, 
The University of Electro-Communication, Chofu, Tokyo 182-8585, Japan}
\address{$^2$Department of Applied Physics, University of Tokyo, Hongo, 
Tokyo 113-8656, Japan}

\begin{abstract}
Based on first principles calculations, 
we study the origin of the large thermopower in Ni-doped LaRhO$_3$ 
and Mg-doped CuRhO$_2$ 
We calculate the band structure and construct the maximally 
localized Wannier functions,
from which a tightbinding Hamiltonian is obtained.
The Seebeck coefficient is calculated within the Boltzmann's equation
approach using this effective Hamiltonian.
For LaRhO$_3$, 
we find that the Seebeck coefficient remains nearly constant within a large
hole concentration range, which is consistent with the experimental 
observation. For CuRhO$_2$, the overall temperature dependence of the 
calculated Seebeck coefficient is in excellent agreement with the 
experiment. The origin of the large thermopower is discussed.
\end{abstract}

\maketitle

\section{Introduction}
The discovery of large thermopower in 
Na$_x$CoO$_2$\cite{terasaki} and the findings in 
cobaltates/cobaltites\cite{4,5,6,7,8} and 
rhodates\cite{9,10} that followed have brought up an 
interesting possibility of finding good thermoelectric 
materials that have relatively low resistivity.
We have recently proposed that the ``pudding-mold'' type band
is the origin for the coexistence of the large thermopower
and the low resistivity in this material. \cite{kuroki}.
Let us first summarize our idea.
Using the Boltzmann's equation, the thermopower is given as 
\begin{equation}
{\bf S}=\frac{1}{eT}{\bf K}_0^{-1}{\bf K}_1
\label{eq1}
\end{equation}
where $e(<0)$ is the electron charge, $T$ is the temperature, 
tensors ${\bf K}_0$ and ${\bf K}_1$ are given by
\begin{equation}
{\bf K}_n=\sum_{\Vec{k}}\tau(\Vec{k})\Vec{v}(\Vec{k})\Vec{v}(\Vec{k})
\left[-\frac{\partial f(\varepsilon)}
{\partial \varepsilon}(\Vec{k})\right]
(\varepsilon(\Vec{k})-\mu)^n.
\label{eq2}
\end{equation}
Here, $\varepsilon(\Vec{k})$ is the band dispersion, 
$\Vec{v}(\Vec{k})=\nabla_{\Vec{k}}\varepsilon(\Vec{k})$ is the 
group velocity, $\tau(\Vec{k})$ is the quasiparticle lifetime,  
$f(\varepsilon)$ is the Fermi distribution function,  
and $\mu$ is the chemical potential. 
Hereafter, we simply refer to $({\bf K}_n)_{xx}$ as $K_n$, and 
$S_{xx}=(1/eT)\dot(K_1/K_0)$ (for diagonal ${\bf K}_0$) as $S$. 
Using $K_0$, conductivity can be given as 
$\sigma_{xx}=e^2K_0\equiv\sigma={1/\rho}$.
Roughly speaking for a constant $\tau$, 
$K_0\sim\Sigma'(v_A^2+v_B^2),
K_1\sim(k_BT)\Sigma'(v_B^2-v_A^2)$
(apart from a constant factor) stand, 
where $\Sigma'$ is a summation over the states in the range of 
$|\varepsilon(\Vec{k})-\mu|<\sim k_BT$, 
and $v_A$ and $v_B$ are typical velocities for the states 
above and below $\mu$, respectively.
If we consider a band 
that has a somewhat flat portion at the top (or the bottom), 
which sharply bends into a highly dispersive portion below (above).
We will refer to this band structure as the 
``pudding mold'' type.
For this type of band with $\mu$ sitting near the 
bending point, $v^2_A\gg v^2_B$ holds for high enough temperature, 
so that the cancellation in $K_1$ is less effective, 
resulting in $|K_1|\sim(k_BT)\Sigma'v_A^2$ and $K_0\sim\Sigma'v_A^2$, 
and thus large $|S|\sim O(k_B/|e|)\sim O(100)\mu$V/K. 
Moreover, the large $v_A$ and  the large FS 
results in a large  $K_0\propto\sigma$ as well, 
being able to give a large power factor $S^2/\rho$, which is important
for device applications.

In the present study, we focus on a possibly related rhodate 
LaRhO$_3$\cite{Shibasaki} with Ni doping and CuRhO$_2$
\cite{kuriyama,shibasaki2} with Mg doping,  
in which large thermopower has been observed.
In Ni doped LaRhO$_3$, the Seebeck coefficient at 300K steeply 
decreases up to the Ni content of $x=0.05$, but then stays 
around $100\mu$V/K up to about $x=0.3$. On the other hand, 
the conductivity monotonically grows, resulting in a 
monotonically increasing power factor (see Fig.\ref{fig2}).
For CuRhO$_2$, the two existing experiments give different results.
In ref.\cite{shibasaki2}, the Seebeck coefficient is found to be 
nearly independent of doping, 
while it decreases with doping in ref.\cite{kuriyama}.

\section{Method}
LaRhO$_3$ has an orthorhombic structure, 
which is distorted to some extent from the ideal cubic 
perovskite structure.
The experimentally determined lattice 
constants are $a = 5.5242(12)$, $b = 5.7005(12)$ and $c = 7.8968(17)$\AA \cite{rene}. 
For comparison, we also calculate the band structure 
for the ideal cubic perovskite structure, where the 
lattice constant is taken as $a = 3.940$\AA.\cite{galasso}
CuRhO$_2$ has an delafossite structure whose 
experimental lattice constants are $a=5.810910$, $c=32.437162$\AA.
We have obtained the band structure of these materials with the 
Quantum-ESPRESSO package\cite{pwscf}. We then construct 
the maximally localized Wannier functions (MLWFs)\cite{MaxLoc} 
for the energy window $-1.75eV < \epsilon_{k}-E_F < -0.64eV$ for 
the ideal structure of LaRhO$_3$, $-1.8eV < \epsilon_{k}-E_F < 0.5eV$ for the distorted structure of LaRhO$_3$ and $-10{\rm eV} < \epsilon_{k}-E_F < 4{\rm eV}$ for CuRhO$_2$, 
where $\epsilon_{k}$ is the eigenenergy of the Bloch states 
and $E_{F}$ the Fermi energy. 
With these effective 
hoppings and on-site energies, the tight-binding Hamiltonian 
is obtained, and finally the Seebeck coefficient 
is calculated using eq.(\ref{eq1}). The doping concentration 
$x$ is assumed to be equal to the hole concentration, and a 
rigid band is assumed.
%
%

\begin{figure}
\includegraphics[width=30pc]{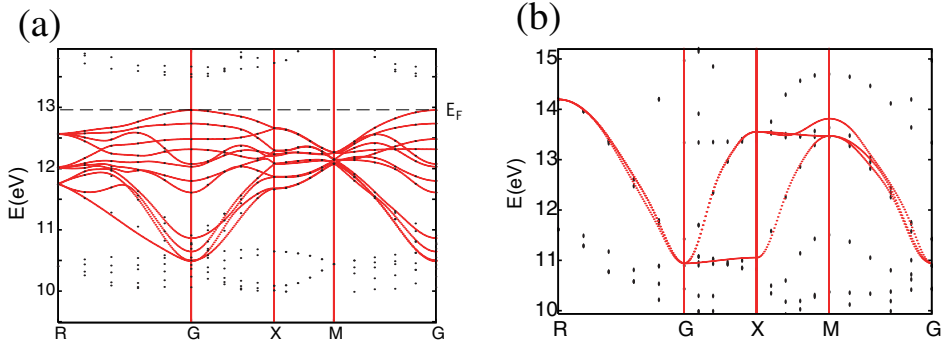}
\caption{\label{fig1}The band structure of the distorted (a) and the 
ideal (b) structure 
the tightbinding mode (solid lines) 
together with the LDA band calculation results (dotted) are shown.}
\end{figure}
\begin{figure}
\includegraphics[width=20pc]{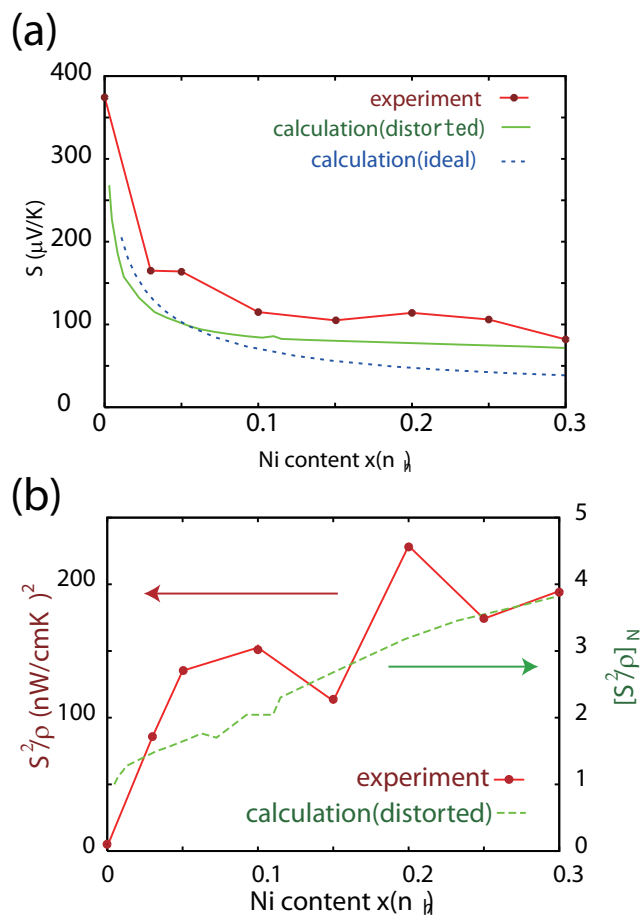}
\caption{\label{fig2}(a) Seebeck coefficient for
the distorted(green) and the ideal(blue) structure. 
The red line is the experimental 
result.\cite{Shibasaki} (b) Power factor of the distorted structure 
normalized at $n_h=0$. The red line is the experimental data.\cite{Shibasaki}}
\end{figure}

\begin{figure}
\includegraphics[width=28pc]{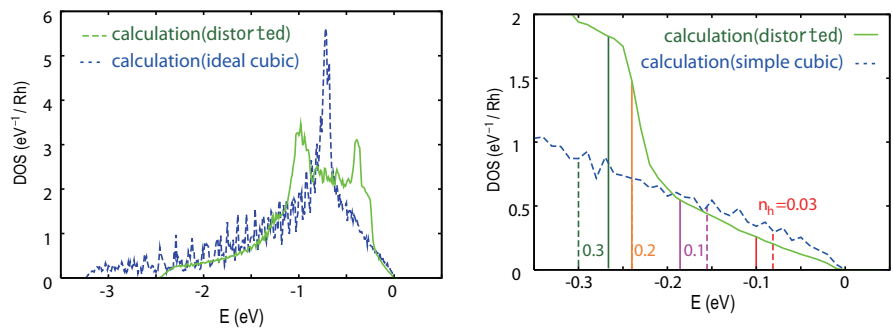}
\caption{\label{fig3}DOS of distorted(green) and ideal(blue) structures. 
The right panel is a blow up of the left.}
\end{figure}

\begin{figure}
\includegraphics[width=20pc]{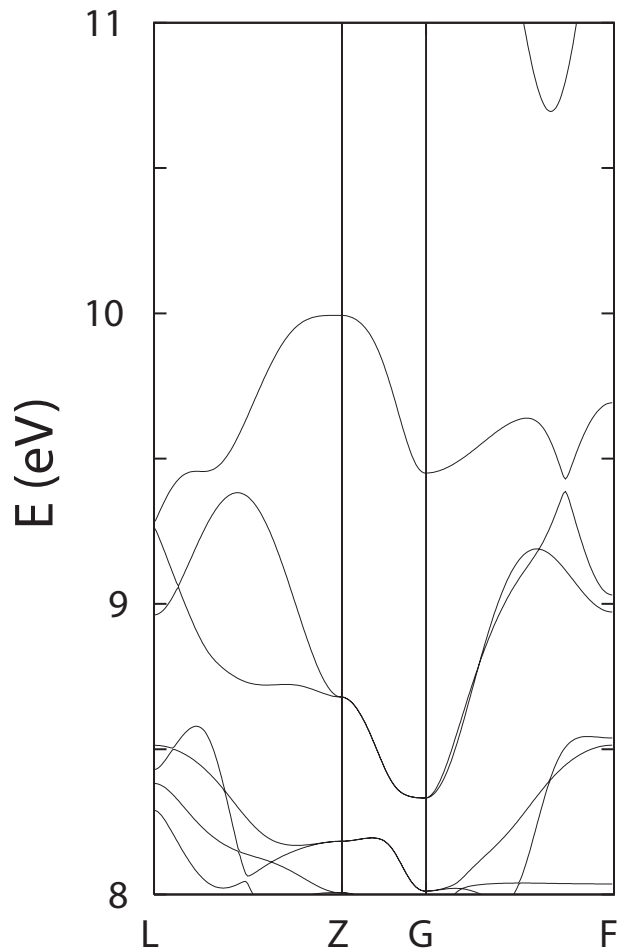}
\caption{\label{fig4} Tightbinding band obtained via maximally localized 
Wannier orbitals }
\end{figure}

\begin{figure}
\includegraphics[width=20pc]{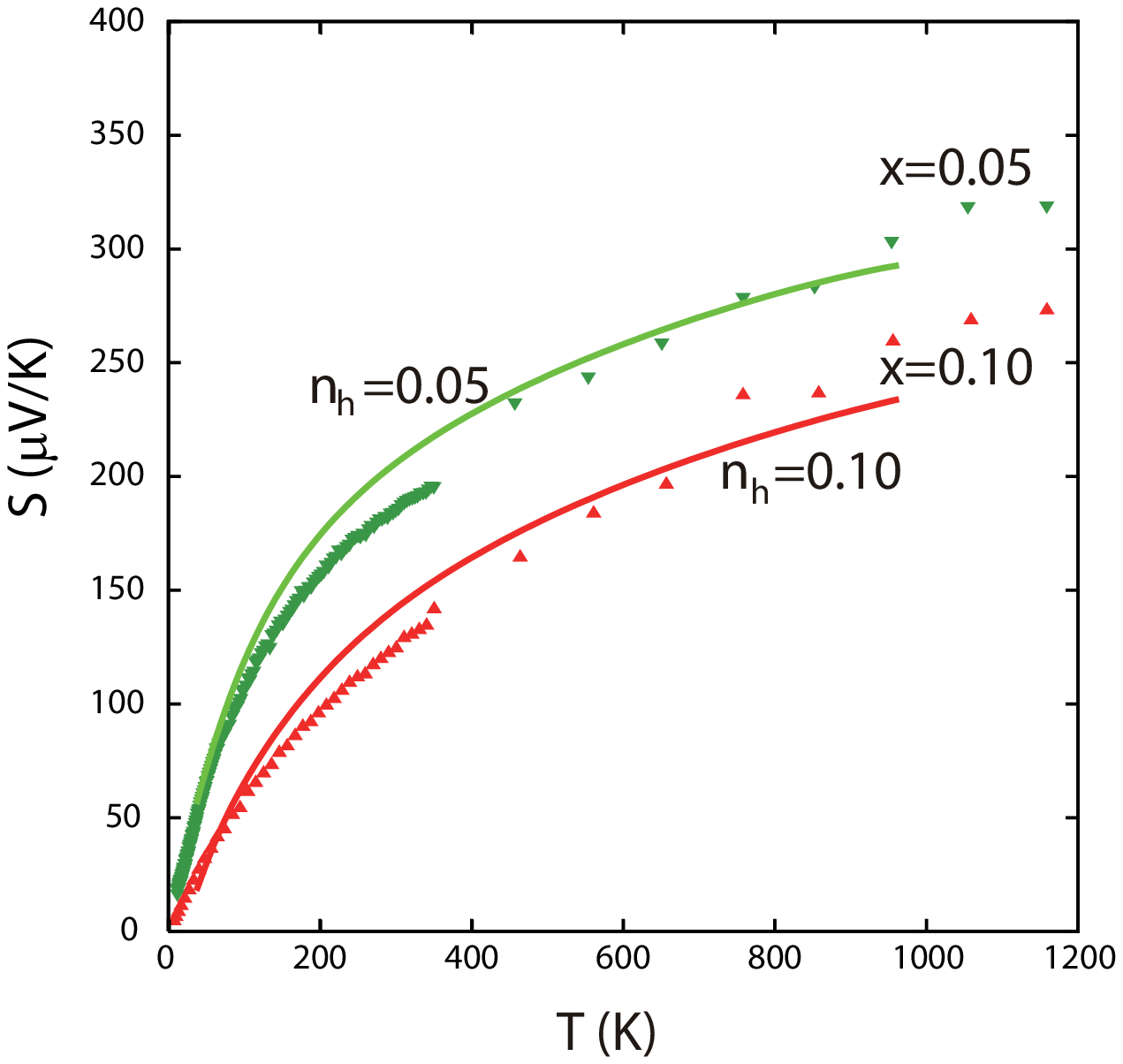}
\caption{\label{fig5} Calculation result of the Seebeck coefficient for 
CuRhO$_2$ with the hole concentration of $n_h=0.05$ and $n_h=0.1$. 
The experimental data are from ref.\cite{kuriyama}}
\end{figure}


\section{Results and Discussions}

We first present results for LaRhO$_3$.
The calculated band structure for the distorted structure of 
LaRhO$_3$ is shown in Fig.\ref{fig1} along with that for the ideal structure.
The tightbinding model Hamiltonian for the distorted structure 
consists of 12 bands (4 Rh per unit cell), while the model for the 
ideal structure contains three $t_{2g}$ bands.
The calculated Seebeck coefficient at 300K 
is shown in Fig.\ref{fig2} as a function of 
hole concentration together 
with the experimental result.\cite{Shibasaki} Here we assume that the 
hole concentration $n_h$ is equal to the Ni content.
It can be seen that the Seebeck coefficient steeply 
decreases with doping with $n_h<0.05$, but stays nearly constant 
for $n_h>0.1$ for the distorted structure in particular. 
As a result the (normalized) power factor monotonically grows with doping, 
which is at least in qualitative agreement with the experimental observation.

Now, in order to understand this peculiar hole concentration dependence of 
the  Seebeck coefficient, we now turn to the density of states (DOS).
The comparison of the DOS between the two structures is shown in 
Fig.\ref{fig3}. The DOS at the band top is 
larger for the ideal case since the three bands are degenerate.
Thus, for low doping, $E_F$ stays closer to the band top for the 
ideal structure, resulting in a larger Seebeck coefficient. 
This is a typical 
example where the multiplicity of the bands 
lead to an enhanced thermopower, i.e., the larger the number of bands, 
the closer the $E_F$ to the band top.
In the case of the distorted structure, as the hole concentration increases, 
$E_F$ lowers and hits the portion of the band with 
a large DOS (Fig.\ref{fig3}left). Therefore, $E_F$ 
hardly moves with doping, resulting in a slow decrease of the 
Seebeck coefficient. 
A large DOS region lies in a lower energy regime 
in the ideal structure, and therefore the Seebeck coefficient 
continues to decrease with doping (up to a larger doping concentration). 
The large Seebeck coefficient of 
about 80$\mu$V/K in the distorted structure can be considered as 
due to the flatness of the top of the bands (around the $\Gamma$ point), 
i.e., the pudding mold type band.

We now move on to CuRhO$_2$. The calculated band structure is 
shown in Fig.\ref{fig4}. Around the $\Gamma$ point, there is again  a 
pudding mold type band, whose top is very flat.
The calculation result of the Seebeck coefficient as a function of 
temperature is shown in Fig.\ref{fig5}. 
We find excellent agreement with the experiment in ref.\cite{kuriyama} 
in a wide temperature range and for the Mg content $x=0.05$ and $x=0.1$.
On the other hand, in ref.\cite{shibasaki2}, the Seebeck coefficient is 
nearly independent on $x$. The origin of the discrepancy between this 
experiment and the present calculation remain as a future problem.

\section{Conclusion} 
To conclude, we have studied the origin of the large thermopower in 
LaRhO$_3$ and CuRhO$_2$. 
From the first principles band calculation results, 
a tightbinding model is obtained via 
the maximally localized Wannier orbitals, and the Seebeck coefficient is 
calculated using the tightbinding model. 
In both materials, the large value of the 
Seebeck coefficient can be considered as 
due to the flatness of the top of the bands 
i.e., the pudding mold type band.
For LaRhO$_3$ in particular, the Seebeck coefficient barely decreases for 
the hole concentration of $n_h>0.1$ in agreement with the experiment, 
which we attribute to the 
peculiar uprise of the DOS near the band top. 

The authors acknowledge I. Terasaki, S. Shibasaki, and M. Nohara for 
valuable discussions and providing us the experimental data. 
Numerical calculation has been done at the Supercomputing Center, ISSP, 
University of Tokyo.
This study has been supported by the Grants in Aid for Scientific 
Research from the MEXT of Japan and from JSPS.

\end{document}